\documentclass[english]{book}
\usepackage{subfigure}
\usepackage{latexsym}
\usepackage{amsfonts}
\usepackage{amssymb}
\usepackage{epsfig}
\usepackage{psfig}
\RequirePackage[dutch,USenglish]{babel}  

\begin{document}


 \pagestyle{empty}
\thispagestyle{empty}
\begin{center}
\Large{Universit\`a degli Studi di Pisa\\ Facolt\`a di Scienze Matematiche, Fisiche e
Naturali\\ Corso di Dottorato in Fisica Applicata}\\

\normalsize{XVII ciclo: 2006}

\vskip 2cm

\Large{Tesi di Dottorato}\\

\bfseries \LARGE{Star formation rate\\in the\\solar neighborhood}
\end{center}
\vskip 3cm
\begin{center}
\begin{tabular}[b]{cc}

\emph{\large{Candidato}} \\
 \mdseries{\large{Michele Cignoni}}\\ 
                                  \\
				   \\
				   \\
				   \\
                                 \emph{\large{Relatore}}\hskip 6cm \emph{\large{Correlatore}}\\
                          \mdseries{\large{Prof. Scilla Degl'Innocenti}}\hskip 3cm \mdseries{\large{Prof. Steven N. Shore}}
\end{tabular}
\end{center}

\clearpage
\frontmatter



\chapter{Summary}
The project, described in this thesis, explores new methods to extract
information through the use of color-magnitude diagrams (CMDs).
In particular, the purpose of this thesis is to provide insight into the star
formation rate in the solar neighborhood, analyzing the observations of the
Hipparcos satellite.

An original technique of comparison has been devised:
\begin{itemize}
\item We employ the Bayesian Richardson-Lucy algorithm to the analysis of
  the observational errors in the CMDs by converting the CMD into an image (in effect, a CMD is an image, the intensity being the number of stars
in a bin of effective temperature and luminosity, affected by a point
spread function that originates from the error distributions of 
the parallaxes and photometry)  and using a
  restoring point spread function derived from the known
  sources of error. The resulting reconstructions should be the best cleaned
  data set with which to perform analyses of the star formation rates;

\item A synthetic population is built via Monte Carlo extractions of masses and
  ages, according the assumed initial mass function (IMF) and the star
  formation rate (SFR). Then, a suitable age-metallicity relation (AMR) gives
  the metallicity. The extracted synthetic stars are placed in the CMD by
  interpolations on the adopted stellar evolution tracks. In order to take
  into account the presence of binary stars, a chosen fraction of stars
  are assumed as binaries and coupled with a companion star.
 Once the number of objects populating the artificial CMD equals that of
 the observed one, the procedure is stopped;

 \item To evaluate the goodness of the assumed model, we transform the
   theoretical and the observational CMDs in two dimensional
histograms, choosing bin sizes in color and in absolute magnitude. Once the
number of theoretical and observational objects is known in each bin, we implement a norma (a function of the residuals, as a $\chi^2$ or
 a Poissonian-$\chi^2$ ) to quantify the differences between the two
   histograms. Then, one searches for the best set of
   parameters in the parameter space (through a simplex
 algorithm). Finally, the confidence
 limit of the results are evaluated through a bootstrap technique;

\item In order to check the sensitivity of the recovered SFR to the different
  parametrical inputs (IMF, binaries, AMR), the algorithm is tested on
  artificial ``Hipparcos'' CMD;

\item After fixing the less important parametrical inputs, the analysis is
  repeated on the real Hipparcos data, previously ``cleaned'' by the
  Richardson-Lucy algorithm. 

\end{itemize}

Brief summary of chapters 1 - 6:\\
Chapter 1 gives an overview on the Galaxy and the solar
neighborhood characteristics. Chapter 2 reviews the statistical basis that will be applied in the
following chapters. Chapter 3 describes the observational data. In chapter 4
we apply the principles of stellar evolution to explain the Hipparcos CMD
morphology. In chapter 5 we examine the qualitative and quantitative
application of the Richardson-Lucy algorithm, in order to obtain an Hipparcos
CMD cleaned from the observational errors.
In chapter 6 we apply the method both to artificial CMDs, showing which
parameters are critical for recovering the star formation rate, and to real
Hipparcos data. In the last sections we test the recovered star formation
against kinematic selection. Finally our results are compared with the ones of
recent papers available in literature.

\mainmatter

\setcounter{page}{1}

%


\section{Conclusions and future work}

The aim of this study was to develop a method for recovering as much
information as possible from binned color-magnitude diagrams (CMDs). In
particular, I applied the method to recover the local SFR from
the Hipparcos color-magnitude diagram. Therefore, I developed a Galactic model
for solar neighborhood: artificial stars are created by a random choice of mass and age from the
assumed IMF and SFR(t), interpolating on a grid of
evolutionary tracks, whose metallicity is determined by the adopted
age-metallicity relation (AMR). A
chosen fraction of these stars are selected as binaries and coupled with
another star randomly chosen with the same procedure.

An artificial CMD is thus generated. The parameter space is searched for the
combination of parameters giving the minimum distance, according a maximum
likelihood statistic,  between the theoretical and the observational CMDs.

In order to reduce the computational time, a set of partial
CMDs was built, using them to produce whatever CMD: each partial CMD was generated
with a step star formation, uniform in a given time interval and zero
elsewhere. Thus, for each combination of IMF, binary distribution and AMR, the
CMD corresponding to any SFR was computed as a linear combination of the partial CMDs.

In order to check the importance of the different parameters (IMF, binaries,
AMR), I tested the algorithm on artificial ``Hipparcos'' CMD (fixing the
minimum luminosity at $M_V\sim 3.5$, the completeness limit of the Hipparcos
sample for stars within 80 pc). At these luminosities, the results indicate
that the recovered SFR is weakly influenced by the right choice of IMF and
binary fraction, but it is largely influenced by the adopted AMR. 
In particular, this result was checked assuming the observational AMR for the
solar neighborhood by N\"ordstrom et al. (2004): in spite of the large
dispersion of this relation, the simulation on the artificial CMD indicate
that most of the information on the underlying SFR is still recovered.
Finally, I applied the algorithm to real Hipparcos data. In contrast with
artificial CMDs, the first problem was the presence of observational
uncertainties (due to photometric and parallax errors).
In order to take into account these uncertainties, I considered an innovative
point of view: a CMD is an image, the intensity being the number of stars
in a bin of effective temperature and luminosity, affected by a point
spread function that originates from the error distributions of 
the parallax and of the photometry.  
Thus I treated the Hipparcos CMD with the same techniques that have been used for image
restoration.  
In practice, I implemented the Richardson-Lucy algorithm to the analysis of color-magnitude diagrams affected
by observational uncertainties: I converted the CMD into an image and, using a restoring
  point spread function function derived from the observation, I
  ``cleaned'' the CMD (taking out the observational errors). 
I showed numerical experiments with artificial CMDs that demonstrate good recovery
  of the original image and establish convergence rates for ideal
  cases with single Gaussian uncertainties and Poisson noise using a
  $\chi^2$ statistic.
Finally, this technique was applied to the Hipparcos sample of the solar
neighborhood, recovering the best ``cleaned'' data set with which to
perform analyses of the local star formation rate.

 Assuming the observational AMR by N\"ordstrom et al. (2004), I tried to
 recover the SFR from this ``cleaned'' CMD. The resulting SFR indicates that
 the recent local history of the Galactic disk was very irregular (a bump
 around 2 Gyr is particularly evident). The mean value increases very steeply
 from 3-4  Gyr ago up to now, in a way qualitatively similar to the findigs of
 Hernandez et al. (2000) and Bertelli \& Nasi (2000). In particular, this
 result is is quite independent against kinematic selections, suggesting that:
\begin{enumerate}
\item The local contamination of
halo and thick disk stars is negligible and/or these populations are older
than 6 Gyr (the possibility to infer the older SFR is
hindered by the completeness limit in absolute magnitude); 
\item In the last 5-6
Gyr, all the stellar generations are well sampled; in other
words, the recovered local SFR is not biased by dynamical diffusion and the
local volume is not ``depleted'' by old disk stars. Moreover, the recovered column-integrated SFR
by Vergely et al. (2002) is very similar to our local SFR, suggesting that
the dynamical diffusion wasn't so efficent in the last 5-6 Gyr. 
\end{enumerate}
The timescale of the recovered SFR seems too long (larger than the dynamical
timescale) to be attributed to local events: an accretion of a satellite
galaxy is suspected. 

This work allowed to develop a general method to extract informations (in our
case the local SFR) from a color-magnitude diagram. The observational CMD is
``cleaned'' with a Richardson-Lucy algorithm, then the chosen information
(SFR) is recovered. In this last process, all the parameters with a not
critical influence are fixed to a given value. However, if the IMF and the binary fraction are
not critical inputs to recover the SFR from Hipparcos stars brighter
than $M_V \sim 3.5$, this condition falls when deeper data are adopted. In
this situation (for example if future possible data from the Gaia mission will
be available), new numerical experiments would be necessary in order to
explore the sensitivity of the result to the adopted parameters. 
The method can be easily applied to the analysis of the SFR of the CMD of
dwarf galaxies, for which the distance among the stars is negligible with
respect to distance of the Galaxy from us.

Another
natural extension of this method could involves the analyses of not-local
Galactic fields: adopting spatial distribution for the different Galactic
components, it would be possible to study the disk SFR on larger scale
lengths. The typical data for a similar analyses would be the color-apparent
magnitude diagram for field observations (stars along the line of sight, with
unknown distance, chemical composition, mass and age). In contrast with local
stars, whose the uncertainties affecting the CMD are mainly due to parallax
errors, for field stars the uncertainties are mainly
photometric: the modified version of the Richardson-Lucy algorithm should
adopt a point spread function, whose the width varies as function of the
magnitude (increasing towards faint magnitudes where the photometric error is
larger). For Galactic field stars the open questions and thus the possible
future research are manifold. For example, comparing the results for local and
field stellar samples could give useful informations. The local sample is the ideal place to
study the disk stars, but it is less informative about thick disk stars. In
contrast, field stars are very informative on the thick disk. The combination
of the two data could give information on the Galaxy as a whole. The local
and the field disk stars could show differences in the AMR, SFR and IMF. In
particular, the possibility that a time dependent IMF is necessary to explain
the data could clear if this function is a real universal quantity.\\\\\\\\\\\\\\\\\

{\bf Complete pdf version is available at:}\\
{\bf  http://etd.adm.unipi.it/theses/available/etd-04032006-122227/}

\backmatter



\end{document}